\documentclass[fleqn,twoside]{article}

\usepackage[headings]{espcrc2}
\usepackage[T1]{fontenc}
\usepackage{amssymb,amsmath,times}
\usepackage{epsfig}
\usepackage{units}
\usepackage{floatflt}
\usepackage{wrapfig}

\readRCS
$Id: espcrc2.tex,v 1.2 2004/02/24 11:22:11 spepping Exp $
\usepackage{graphicx}
\usepackage[figuresright]{rotating}

\title{New p+C data in fixed target experiments and the muon component in extensive air showers}

\author{C. Meurer\address[FZK]{Institut für Kernphysik, Forschungszentrum Karlsruhe GmbH,
        Postfach 3640, 76021 Karlsruhe, Germany},
        J. Blümer\addressmark[FZK], R. Engel\addressmark[FZK], A. Haungs\addressmark[FZK], M. Roth\addressmark[FZK] 
        and the HARP collaboration \thanks{See author list of \cite{HARPpAl}}}

\runtitle{New p+C data in fixed target experiments and the muon component in EAS}
\runauthor{Christine Meurer}

\begin{document}

\begin{abstract}
The interpretation of extensive air shower (EAS) measurements is strongly dependent on the hadronic interaction models used for simulating reference showers. We study the importance of low-energy hadronic interactions in simulated air showers, generated with the simulation package CORSIKA, for the observed characteristics of extensive air showers. In particular we investigate in detail the energy and the phase space regions of secondary particle production which are most important for muon production.
This phase space region is covered by fixed target experiments at CERN. In the second part of this work we present preliminary momentum spectra of secondary $\pi^{+}$ and $\pi^{-}$ in p+C collisions at \unit[12]{GeV/c} measured with the HARP spectrometer at the PS accelerator at CERN. In addition we use the new p+C NA49 data at \unit[158]{GeV/c} to check the reliability of hadronic interaction models for muon production in EAS.
Finally, possibilities to measure relevant quantities of hadron production 
in existing and planned accelerator experiments are discussed.

\vspace{1pc}
\end{abstract}

\maketitle

\section{Relation of muons in EAS to hadronic interactions}

In order to extract information (energy and particle type) on the primary particle from ground based air shower measurements, simulation of air showers using electromagnetic and hadronic interaction models are necessary. These simulations show uncertainties which come mainly from hadronic interaction models.
Muons measured at ground level are one of the main ingredients to infer energy and mass of the primary particle of an air shower. Furthermore muons are sensitive to the characteristics of hadronic interactions. 

\begin{figure}[h!]
\centering
\includegraphics[width=0.45\textwidth, bb=  10 20 511 345,clip]{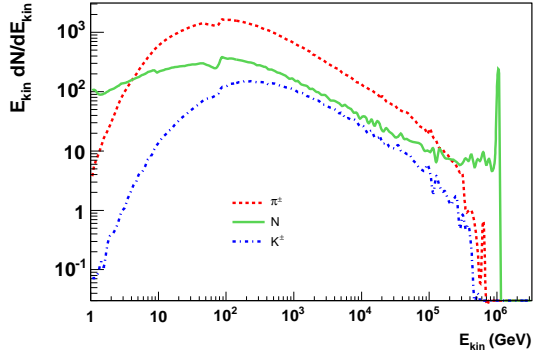}
\vspace{-0.5cm}
\caption{\label{Edis_gm} Energy distribution of {\it grandmother particles} in vertical proton showers with a primary energy of \unit[$10^{15}$]{eV} simulated with CORSIKA (GHEISHA and QGSJET-01 implemented)\cite{C2CR}. \vspace{-0.5cm}}
\end{figure}

One aim of this work is to specify the low-energy hadronic interactions which are important for the muon production in EAS. We simulate extensive air showers with a modified version of the simulation package CORSIKA \cite{CORSIKA}. This special version stores the information of the {\it last hadronic interaction} where mesons are produced which decay into muons. The initiator hadron of the {\it last hadronic interaction} is called the {\it grandmother particle} and the meson which decays into a muon is named {\it mother particle}. For the following analysis vertical proton showers with a primary energy of $10^{15}$eV are used. In this CORSIKA simulation, the hadronic interaction model GHEISHA \cite{GHEISHA} is applied for energies up to \unit[80]{GeV} and QGSJET-01 \cite{QGSJET} for higher energies. Simulations with other models are in preparation. 
In Fig.~\ref{Edis_gm} the energy spectra of {\it grandmother particles} of different particle types are shown. Most of the {\it grandmother particles} are pions but also about 20\% are nucleons and only a few are kaons. The most important energy region reaches from \unit[8]{GeV} up to \unit[1000]{GeV}. The  transition from the low to the high-energy model is visible in all spectra as a step at \unit[80]{GeV}. The largest fraction of {\it mother particles} are pions with around 90\%, the rest are mostly kaons. For more details see \cite{C2CR}.

\section{Comparison: EAS and fixed target experiment}
In particular the forward hemisphere of the phase space of secondary particles is important for muon production in EAS as shown in detail in chapter~\ref{phs}. Therefore fixed target experiments are well suited to measure reactions comparable to the hadronic interactions in EAS. In contrast to collider experiments, in fixed target experiments the forward direction of the phase space of secondary particles is accessible and nuclear targets can be used. 
Comparing the {\it last} interaction in EAS with collisions studied at accelerators, one has to keep in mind that the {\it grandmother particle} corresponds to the beam particle and the {\it mother particle} is equivalent to a secondary particle produced in e.g.\ a minimum bias p-N interaction. 
The most probable energy of the {\it grandmother particle} is within the range of beam energies of fixed target experiments e.g. at the PS and SPS accelerators at CERN.

\section{Relevant phase space regions}\label{phs}
\begin{figure}[h!]
\begin{minipage}{0.42\textwidth}
\centering
\includegraphics[width=0.93\textwidth, bb=  10 20 520 505,clip]{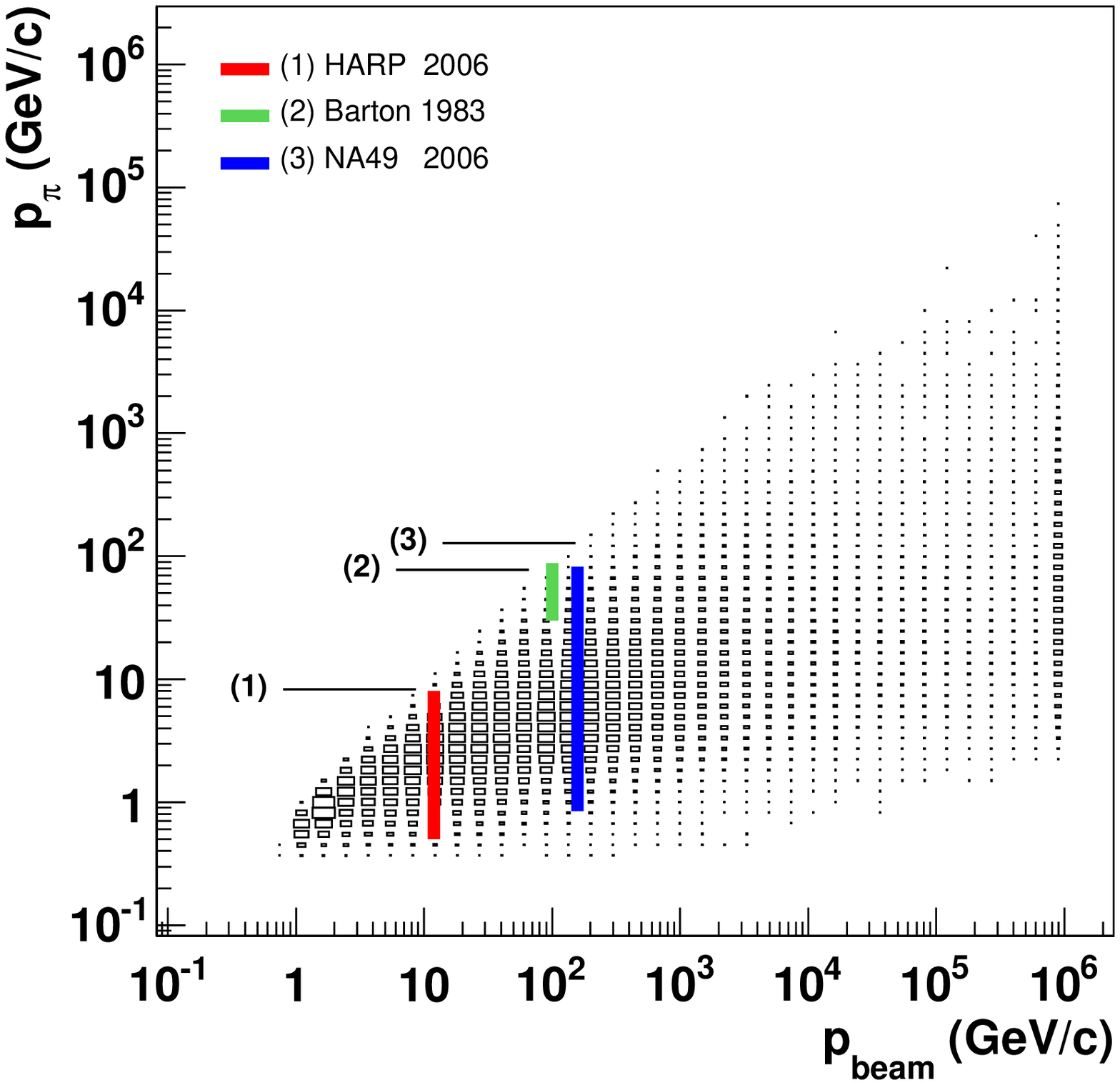}
\includegraphics[width=0.93\textwidth, bb=  10 20 520 505,clip]{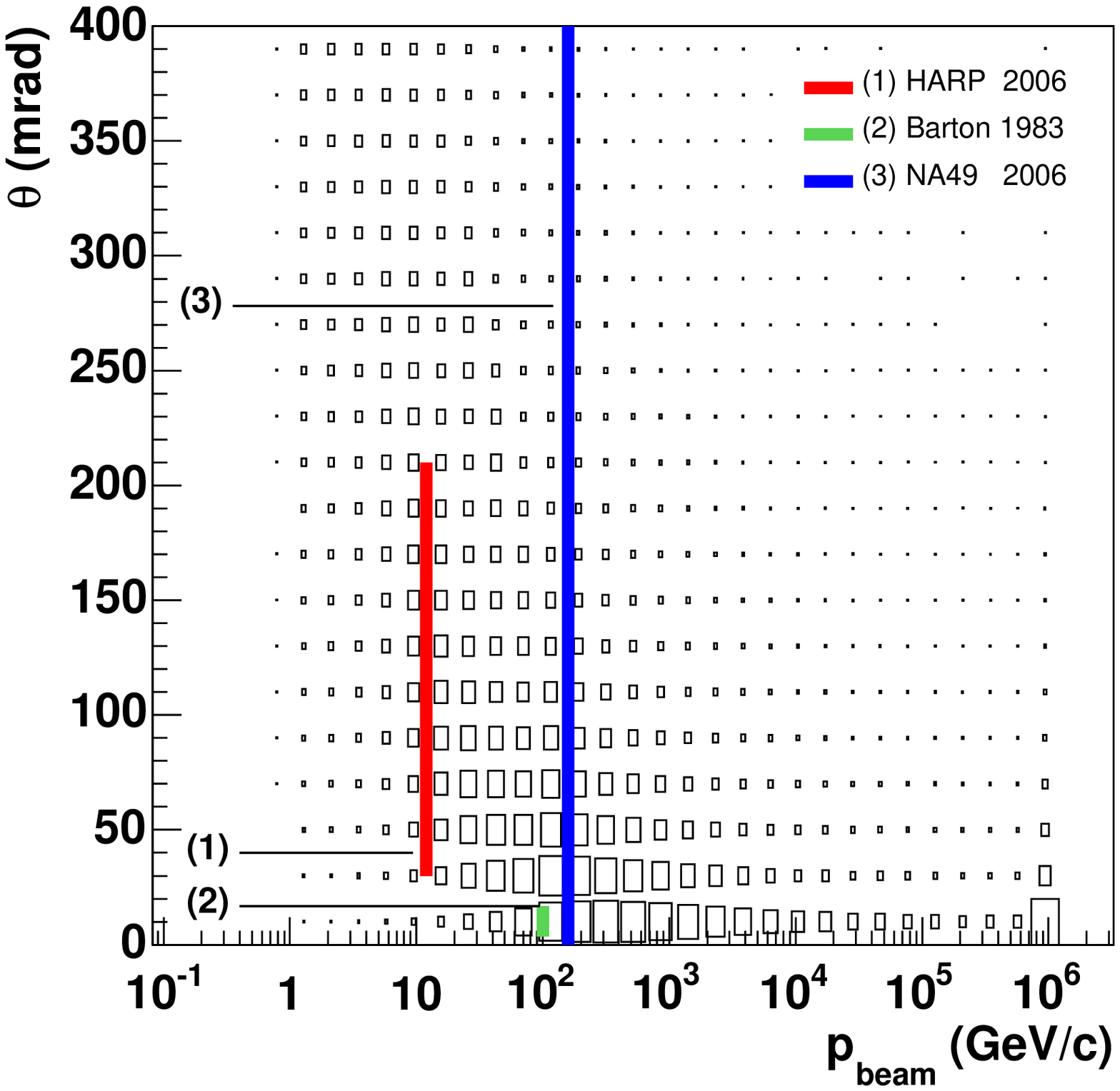}
\end{minipage}
\vspace*{-0.5cm}
\caption{\label{datasum_pC} Coverage of the phase space regions of relevance to EAS (box histograms) by existing fixed target data using a proton beam and a carbon target (shaded/colored regions).  Upper panel: total momentum of secondary pions vs. total momentum of proton projectiles. Lower panel: angle between beam and secondary particle momentum vs. beam momentum. \vspace*{-0.5cm}}
\end{figure}

Motivated by the availability of protons as beam particles at accelerators we consider only those {\it last} interactions in EAS that are initiated by nucleons. The phase space regions of {\it mother particles} produced in the {\it last} interaction in EAS are shown in Fig.~\ref{datasum_pC}. The relevant phase space is given as box histogram. The size of the boxes indicates the relative importance of the phase space region. The momentum of the {\it grandmother particle} is given as abscissa and the {\it mother particle} observable as ordinate. In both figures, existing fixed target p+C data are indicated by shaded (colored) regions. In the past the only measurement of p+C collisions, which was not limited to a fixed angle, was the experiment done by Barton et al. \cite{Barton83}. These data were collected using the Fermilab Single Arm Spectrometer facility in the M6E beam line. A proton beam with a beam momentum of \unit[100]{GeV/c} and a thin carbon target (\unit[1.37]{g cm$^{-2}$}) was used. The phase space of the secondary particles (pions, kaons, protons) covers only a small part of the phase space of interest to EAS.

New p+C data at \unit[12]{GeV/c} and \unit[158]{GeV/c}, taken by the CERN experiments HARP \cite{HARP} at the PS accelerator and NA49 \cite{NA49NIM} at the SPS accelerator, are available now. 
At both beam momenta the secondary particles ($\pi^{+},\pi^{-}$) are measured in a broad momentum region and up to large angles. At \unit[12]{GeV/c}, the HARP data cover secondary momenta from \unit[0.5]{GeV/c} to \unit[8]{GeV/c} and an angular range from \unit[30]{mrad} to \unit[210]{mrad}. 
The NA49 data are taken in the momenta range \unit[0.85]{GeV/c} $\lesssim p \lesssim $ \unit[82.6]{GeV/c} and at angles up to \unit[440]{mrad} \cite{NA49pC}. 
In addition there exist also several p+Be data sets (see references in \cite{C2CR}). However, as simulations with hadronic interaction models show, the particle production in p+air collisions is much more similar to p+C collisions than to p+Be reactions because of the smaller difference in the atomic mass.

\section{Data from fixed target measurements}

\subsection{NA49 p+C data at 158 GeV/c}
\begin{figure}[h!]
\begin{minipage}{0.44\textwidth}
       \centering
       \includegraphics[width=0.93\textwidth, bb=  35 20 520 373,clip]{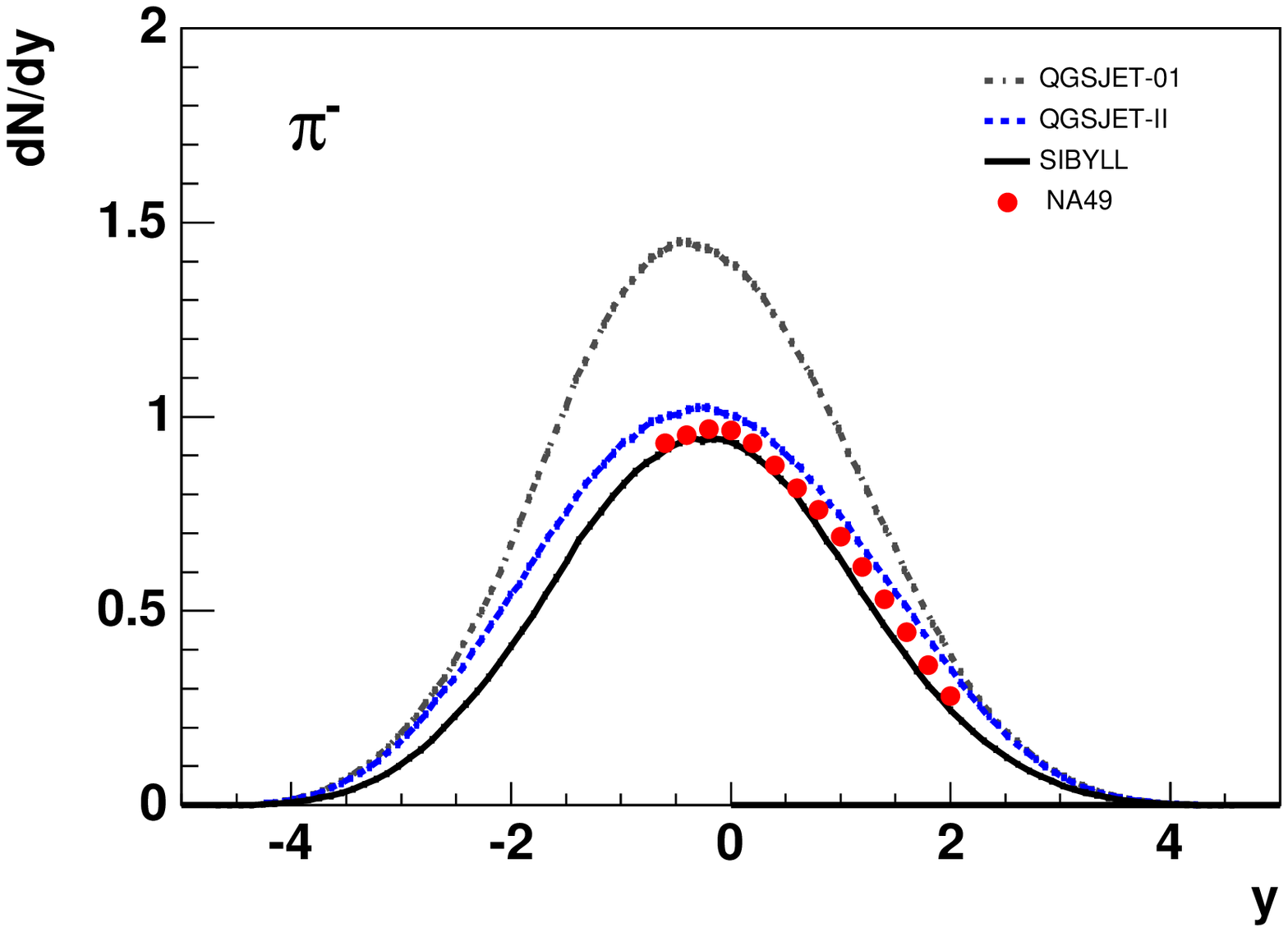}
       \includegraphics[width=0.93\textwidth, bb=  35 20 520 373,clip]{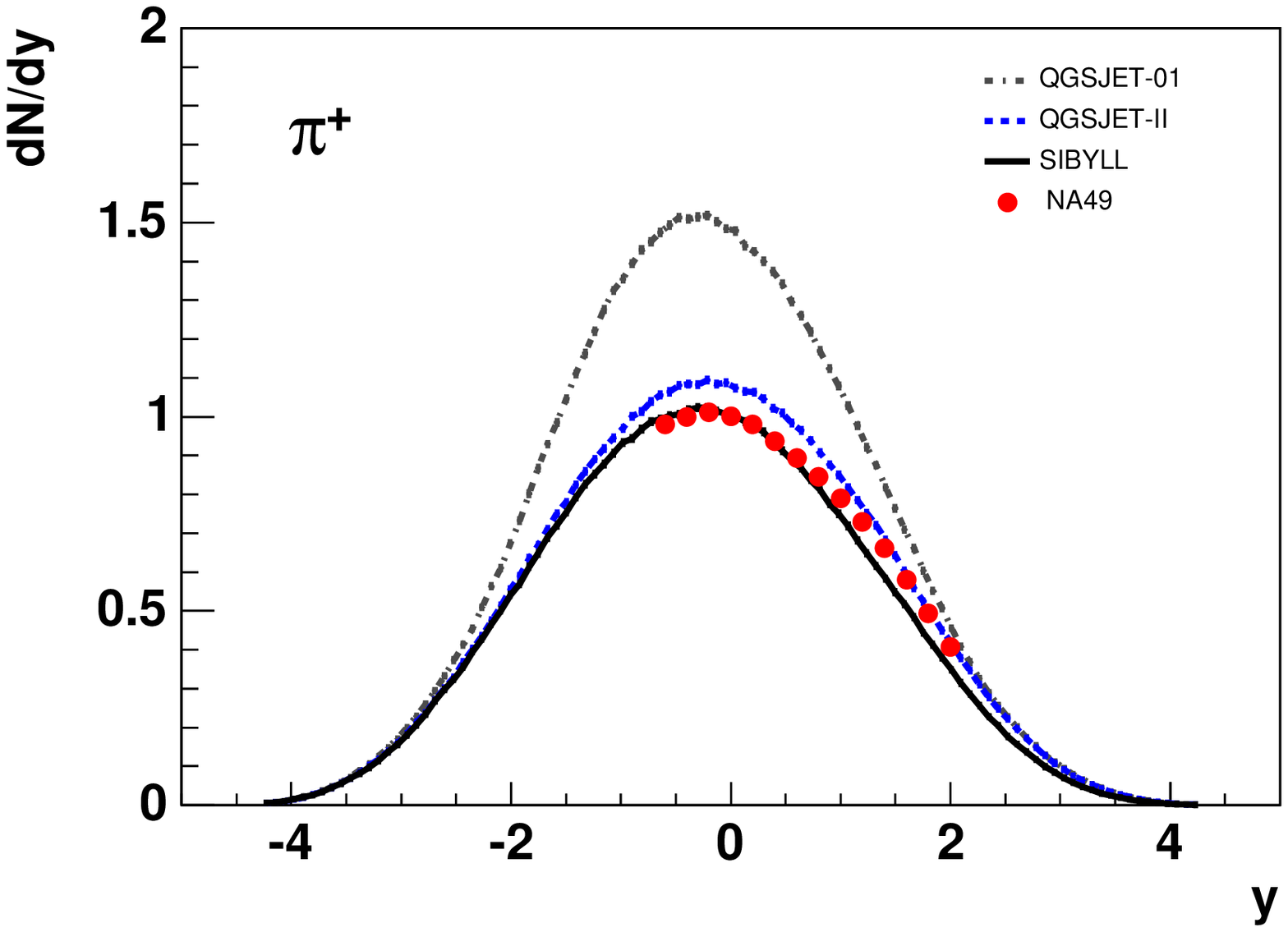}
\end{minipage}
\vspace*{-0.5cm}
\caption{Rapidity spectra of pions produced in p+C collisions at \unit[158]{GeV/c}. The filled circles indicate the NA49 measurements \cite{NA49pC}. The three lines show the results of QGSJET-01 \cite{QGSJET}, QGSJET-II \cite{QGSJET21,QGSJET22} and SIBYLL2.1 \cite{SIBYLL1,SIBYLL2} simulations. \label{NA49pCrap_pi} \vspace*{-0.5cm}}
\end{figure}

Pion spectra produced in p+C collisions with a beam momentum of \unit[158]{GeV/c} measured by NA49 became available this summer \cite{NA49pC}. In Fig.\ref{NA49pCrap_pi} rapidity spectra of secondary pions are compared with simulations done with QGSJET-01, QGSJET-II \cite{QGSJET21,QGSJET22} and SIBYLL2.1 \cite{SIBYLL1,SIBYLL2}. SIBYLL2.1 and QGSJET-II predictions show a reasonable agreement with the data, but QGSJET-01 overestimates the measured spectra by a factor $\sim$1.5 in the central rapidity region. Comparing QGSJET-01 and QGSJET-II concerning the shape of the spectra, QGSJET-II predicts harder pion spectra than the older version.

\subsection{HARP p+C data at 12 GeV/c}
In these proceedings, we present for the first time momentum spectra of secondary pions in p+C collisions with a beam momentum of \unit[12]{GeV/c} measured by the HARP spectrometer \cite{HARP}. The thickness of the carbon target is equivalent to 5\% nuclear interaction length (\unit[3.56]{g/cm$^{2}$}). For the selection of the secondary particles ($\pi^{+},\pi^{-}$) in the forward hemisphere four walls of drift chambers are used. The separation of particle types is done with different detector components:
The time-of-flight measurement allows pion--kaon and pion--proton separation to be performed up to \unit[3]{GeV/c} and beyond \unit[5]{GeV/c}, respectively. Additional a Cherenkov detector is used to separate pions from protons and kaons above \unit[2.5]{GeV/c} \cite{HARPpAl}. 

\begin{figure*}[t]
\centering
\includegraphics[width=0.95\textwidth, bb=  1 10 555 350,clip]{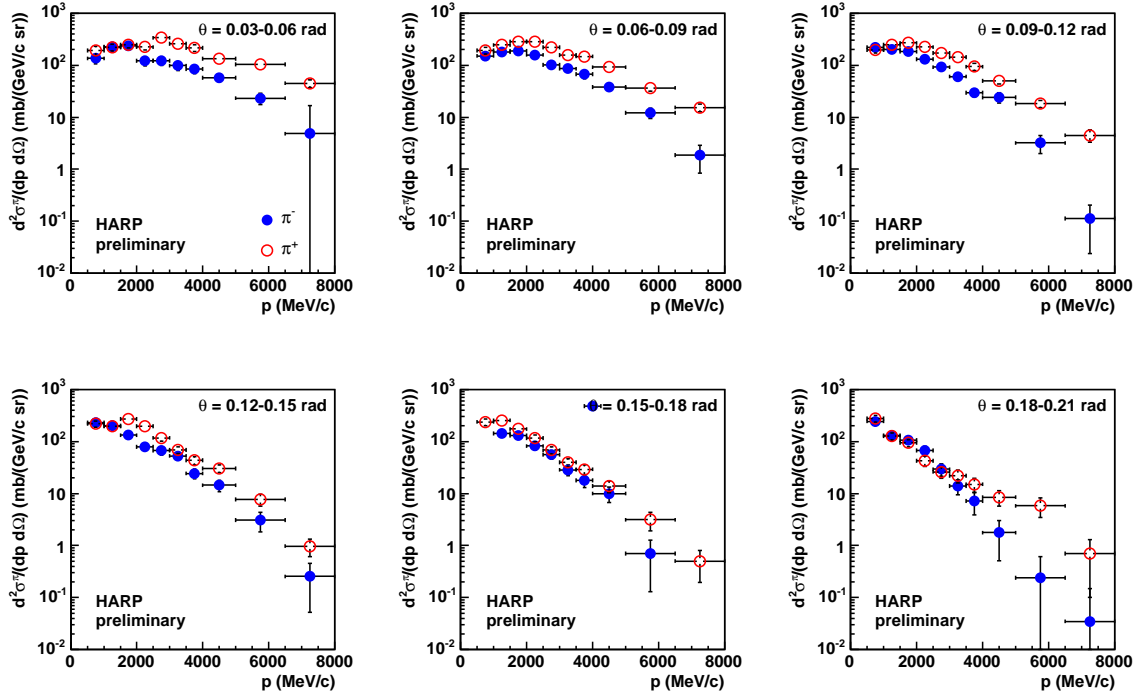}
\vspace*{-0.1cm}
\parbox{0.95\textwidth}{\caption{Preliminary momentum spectra of secondary $\pi^{-}$ (filled circles) and $\pi^{+}$ (open circles) mesons in p+C reactions at \unit[12]{GeV/c} measured with the HARP spectrometer at the PS accelerator at CERN. The six different panels show the spectra in different angular bins from \unit[0.03]{rad} to \unit[0.21]{rad}.\label{pC12GeV_pi}}} 
\end{figure*}

In Fig.\ref{pC12GeV_pi} the preliminary $\pi^{+}$ and $\pi^{-}$ momentum spectra are presented in six different angular bins starting from \unit[0.03]{rad} up to \unit[0.21]{rad}. The cross section of the $\pi^{+}$ production is consistently higher for all angular bins than the cross section of $\pi^{-}$, especially in the higher momentum regions. This behaviour can be understood in terms of the leading particle effect which influences mainly positive pions. The error bars indicate the statistical and systematic errors. The size of both errors are of the same order of around 4 to 14\% depending on the secondary momentum. Only for the highest momenta the statistical error of the $\pi^{-}$ amounts more than 20\%. Generally the statistical errors of the $\pi^{-}$ are some what larger than that for the $\pi^{+}$. The dominant contributions to the systematic error are tertiary subtraction and momentum scale.

\section{Conclusions and outlook}
New p+C data from NA49 and HARP are available now. They cover a broad range of phase space which is important for the EAS. However, only 20\% of the {\it grandmother particles} in EAS are protons and most are pions. 
Currently an analysis of HARP $\pi$+C data is in progress to help tuning hadronic interaction models for primary pions. 
Future data of the recently proposed upgraded MIPP experiment at Fermilab \cite{MIPP1,MIPP2} of p+C, $\pi$+C and $K$+C collisions at 20, 60 and \unit[120]{GeV/c} would give important input for hadronic interaction models at higher energies. As a future project, an upgrade of the NA49 detector is planned \cite{NA49future}. This will allow the measurement of reactions on p+C and $\pi$+C at the energies of 30, 40, 50 and \unit[158]{GeV}.

\end{document}